\magnification \magstep 1
\baselineskip 19 pt
\noindent 
\bf \centerline{The Dynamical and Static  Casimir Effects } 
\centerline{and} 
\centerline{the Thermodynamic Instability}
\rm
\bigskip
\bigskip
\centerline{E.~Sassaroli$^{*}$ [1, 2], 
Y.~N.~Srivastava [3,2],  
J.~Swain [2] and A.~Widom [2] }
\centerline{1. Center for Theoretical Physics, Laboratory for 
Nuclear Science 
and Department of Physics}
\centerline{Massachusetts Institute of Technology, Cambridge, MA 02139, USA}
\centerline{2. Physics Department, Northeastern University Boston, 
Massachusetts, 02115, USA}
\centerline{3. Dipartimento di Fisica and I.N.F.N, Universita' degli 
Studi di Perugia, 06100 Perugia, Italy}
\bigskip
\medskip
\centerline{(Talk presented by E. Sassaroli at the 
Casimir effect topical group meeting }
\centerline{ITAMP, Smithsonian-Harvard Center for
Astrophysics, March 1998)} 
\bigskip
\medskip
\centerline{\bf ABSTRACT}
\bigskip
\rm  \par
\noindent The dynamical Casimir effect 
for the ideal case of two perfectly conducting non-charged parallel plates,
is discussed using the zero-point energy summation method  
to the first order in perturbation theory.
We show that it is possible
to create photon radiation when the two plates are modulated rapidly 
in time.
Moreover we point out that the static Casimir energy 
between two conducting  non-charged parallel plates violates 
the thermodynamic stability
condition normally associated with the second law of thermodynamics.
\vskip .8in\
$$
$$

\par
\noindent $^*$This work is supported in part by funds provided by 
the U. S. Department
of energy (D.~O.~E) under cooperative research agreement 
$\#$DF-FC02-94ER40818
\vfill
\eject
\par
\noindent 
\bf 1. Introduction
\bigskip
\rm
It has become common practice to use the label  Casimir effect 
[1] whenever
one wishes to describe a non-classical electromagnetic force of
attraction between two perfectly conducting non-charged parallel plates.
The central idea behind this effect is that the vacuum can 
be ``perturbed'' by
inserting objects to produce measurable effects. The electromagnetic
vacuum near a conducting plate is different from the one in the empty space.
The perturbed vacuum gives rise to an attractive long range force between 
the two plates. 

Over the past fifty years, Casimir forces have been 
calculated for a variety of different geometries. Examples include plane 
dielectric surfaces, cavities, polarizable particles, etc. 
Different methods have been developed to evaluate these forces: the
mode summation method, the local Green function method, source 
theory, etc. [2-5]. 

Regardless of the way the Casimir forces are calculated, they
are obtained by holding the geometry either fixed or by varying it 
only quasi-statically. We use the term {\it dynamical Casimir effect} [6] 
when
the geometry of the system is varied more quickly in time. In this case 
photon production becomes possible simply by {\it accelerating the vacuum}.

In this paper we discuss the dynamical Casimir effect for the ideal case 
of two
perfectly conducting non-charged parallel plates [7]. The 
frequencies of the
zero-point electromagnetic energy between the plates depend on the 
distance between them. If we modulate the distance in time the 
zero-point electromagnetic
energy becomes also a function of time. The result is the possibility of 
photon production, if the modulation occurs quickly enough in time.
     
There is much recent discussion in the literature about the possibility
of photon production by the dynamical Casimir effect, see for example 
[8-24] and references there.

Moreover, we point out here that   
the Casimir vacuum between two perfectly conducting plates is 
thermodynamically 
instable having a zero temperature negative compressibility [25]. 
$$
$$
\bigskip
\par  
\noindent
\bf 2. The dynamical Casimir effect
\bigskip
\rm \par
The static Casimir energy for two perfectly conducting plates separated
by a distance {\it d} 
{\it ( L$_{x}$ = L$_{y}$ = L} and {\it L$_{z}$ = d} with {\it L $\gg$ d )}
is given by the difference between the zero-point energy when the plates
are placed at a distance {\it d} and when the distance between them is
infinite 
$$
U(d) = E_o(d) - E_{o}(\infty) = -{\pi^2 \hbar c L^2\over 720 d^3}.
\eqno(1)
$$ 

Now suppose that the distance {\it d} varies as a function of time {\it d(t)}. 
We assume a constant initial value $d_i= a$
and after a time $\Delta t$ the modulation stops and the distance {\it d} 
reaches a final value, which for simplicity we assume equal to the  
initial value $d_f=a$. 

The possibility of photon production due to the effect of the plate 
modulation can be described as follows.\par

The Hamiltonian for a single electromagnetic mode before the modulation is
$$
H_{in}=\hbar \omega_{in}(a^{\dagger}_{in}a_{in}+ {1\over 2}),
\eqno(2)
$$
with 
$$
\omega_{in} = \omega =  c\ \sqrt{\left ({\pi l\over L}\right )^2 + \left
({\pi m\over L}\right )^2 +\left ({\pi n\over a}\right )^2},
\eqno(3)
$$
where l, m, n = 0, 1, 2, $\ldots$, with the restriction that only one  
integer at a time can be zero.
After the modulation the Hamiltonian is
$$
H_{out}=\hbar \omega_{out}(a^{\dagger}_{out}a_{out}+ {1\over 2}),
\eqno(4)
$$
with $\omega_{out}=\omega_{in}$.

We can relate the ladder operators $a_{in}$ and $a_{out}$ through a Bogolubov 
transformation
$$
a_{out}=Ua_{in}+Va^{\dagger}_{in}, \ \ \ \ \ 
a^{\dagger}_{out}=U^{*}a^{\dagger}_{out}+V^{*}a_{out},
\eqno(5)
$$
where the relation 
$$
|U|^2-|V|^2=1,
\eqno(6)
$$
is required by the equal time commutation relation.

The mean number of photons produced through  the modulation 
of the single electromagnetic mode [26]  
$$
N=<0|a^{\dagger}_{out}a_{out}|0>,
\eqno(7)
$$  
obeys the relation 
$$
N=|V|^2,
\eqno(8)
$$
through Eqs. (5) and (6), with $a_{in}|0>=0$.

To the first order in perturbation theory $N$ can be determined in terms
of the scattering solutions of the equation [27] 
$$
{d^2 \over dt^2}q(t)+\omega^2(t)q(t)=0,
\eqno(9)
$$
with 
$$
\omega (t) =  c\ \sqrt{\left ({\pi l\over L}\right )^2 + \left
({\pi m\over L}\right )^2 +\left ({\pi n\over d(t)}\right )^2},
\eqno(10)
$$
subject to the boundary condition that for large {\it t} the time dependent 
frequency $\omega (t)$ goes to the unperturbed values  
$$
\omega(t \rightarrow  -\infty)=\omega_{in}=\omega, \ \ \ \  \
\omega(t \rightarrow  \infty)=\omega_{out}=\omega.
\eqno(11)
$$
Eq. (9)  is formally identical to a  one-dimensional Schr$\rm \ddot o$dinger
equation, where t $\rightarrow$ x. \par 
If the potential $\omega (t)$ is symmetrical respect to the origin and goes
to zero  for large values of $|t|$, then the asymptotic solutions of 
the Eq. (9) are
$$
q(t) = \cases{Ae^{i\omega t}+Be^{-i\omega t}  &$t\rightarrow -\infty$,\cr
			  Fe^{i\omega t}+Ge^{-i\omega t},
&$t\rightarrow +\infty$.\cr} 
\eqno(12)
$$

We can relate the coefficients {\it A, B, F, G} through the matrix 
M defined by
$$
\left ( \matrix {A \cr B} \right )= 
\left( \matrix {\alpha_1 +i\beta_1 & i\beta_2 \cr
	       -i\beta_2 & \alpha_1-i\beta_2\cr}\right)
\left ( \matrix {F \cr G} \right ),
\eqno(13)
$$
where $\alpha_1$,  $\beta_{1,2}$ are real numbers satisfying the relation
$$
\alpha_1^2+\beta_1^2-\beta_2^2=1.
\eqno(14)
$$

It is possible to express N in terms of the reflection coefficient $R$ 
$$
R={|B|^2 \over |A|^2}
\eqno(15)
$$ 
as 
$$
N=|V|^2={R\over 1-R}.
\eqno(16)
$$
Eq. (16) can be easily obtained by noticing that $[1/(1-R)-R/(1-R)]$ 
satisfies Eq. (6) with $|U|^2=1/(1-R)$.

Suppose now that the distance between the two plates in modulated 
periodically in time during some time interval $\Delta t$ by a succession of   
pulses. Each pulse satisfying Eqs. (9) and (10).
The number of photons produced in a given electromagnetic mode is [7]
$$
N=\beta^2_2\left ( {sin(r \gamma) \over sin(\gamma)} \right )^2,
\eqno(17)
$$
where $r$ is the number of pulses and $\gamma$ satisfies the condition
$$
cos(\gamma T)=\alpha_1 cos(\omega T)+\beta_1 sin(\omega T),
\eqno(18)
$$
with $T$ modulation period.

The total number of photons produced is obtained by taking 
the sum over all the electromagnetic modes up to a cut-off 
frequency $\nu_c \sim 10^{15}$ Hz.
$$
 N_{tot} = \sum_{\bf k\rm \lambda} N_{\bf k \rm \lambda}(\omega),
\eqno(19)
$$
where $\lambda$ = 1, 2 is the polarization index.

We can evaluate $N_{tot}$ for the following interesting case. For 
$L \gg a$ the frequency $\omega$ in Eq. (10) can be written as 
$$
\omega=2 \pi \nu={\pi c \over a}\sqrt{x^2 +n^2}, \ \ \  x=ak_{||}/\pi,
\eqno(20)
$$
with $k^2_{||}=k^2_x +k^2_y$. If $a \sim 10^{-4}$ cm the frequency 
$\omega \sim 10^{15}$ Hz i.e. in the visible region of the 
electromagnetic spectrum. 

It is possible
to show that for the value of $a \sim 10^{-4}$ cm and for a modulation 
period $T \sim 10^{-3}$ s the total number of photons produced 
per unit area and per pulse in the limit
$r \rightarrow \infty$  is [7]
$$
{\cal N}\sim {1\over a^2}\sum_{n=1}^{n_c}\int{dxx\beta_2^2(x,..)}.
\eqno(21)
$$
The dots in Eq. (20) means that the function $\beta_2$ is not only
a function of $x$ but also of the parameters which characterize each pulse,
such as for example  the duration and the modulation strength. If for 
those values the term $1/a^2$ gives the dominant contribution  
in Eq.(21) then the photon production can be appreciable, as it can
be seen through the following examples.
$$
$$
\par
\noindent \bf 3. Examples
\bigskip
\rm
As a first example, we consider a succession of time pulses such that for each pulse 
the distance between the two plates varies as  
$$
d^2(t) = { a^2\over
1 + {b^2\over \cosh^2(\pi t/\tau)}},
\eqno(22)
$$
where $\tau$ describes the pulse duration and b determines the 
modulation strength. We assume $b \sim 5$, $a$ and $T$ given above. 

Therefore  the modulated single mode 
frequency is
$$
\omega^2(t)  = \omega^2+
 \left ({c \pi  nb\over a\ cosh(\pi t/\tau)}\right )^2,
\eqno(23)
$$
where $\omega$ is given by Eq. (20).
The function $\beta_2$ for this case is 
$$
\beta_2\sim {1\over sinh^2(\omega \tau)}. 
$$
The function $\beta_2$ goes to zero very quickly and therefore also the 
integral in Eq. (21) except when 
$\tau \sim 1/\omega$. Therefore the photon production can be appreciable only 
for very sharp pulses. 
\medskip
As a second example we consider a succession of rectangular potential
barriers. In the valleys of the potential $\omega (t)$ the distance between the two plates is  
$$
d^2(t)=a^2,
\eqno(24)
$$
and in the hills of the potential $\omega (t)$
$$
d^2(t)={a^2\over 1+b^2}
\eqno(25)
$$
where b is the modulation strength. \par
The function $\beta_2$ is given by
$$
\beta_2={1\over 2}\left ( 
{\omega \over \omega_2}- {\omega_2 \over \omega } \right )
sin(2\omega_2T),
\eqno(26)
$$
where $\omega_2$ is 
$$
\omega_2={c \pi \over a}\sqrt{x^2+n^2(1+b^2)}.
\eqno(27)
$$

For the values of $a$, $T$ and $b$ chosen above, the
estimated  number of produced photons is
$$
\sim 10^{11}/{\rm area-pulse}, \eqno(28)
$$
mainly in the visible region.

Therefore if there is a rapid modulation of the plates photon production
becomes possible.

The dynamical Casimir effect has been considered as  possible 
explanation of the phenomenon of sonoluminescence [6].
In the sonoluminescence experiments air bubbles in water, under the effect 
of sound waves, expand up to the dimension  $\sim$ 50 $\mu$m and 
then collapse very quickly to a dimension $\sim$ 0.5 $\mu$m. During the  
collapse a flash of light of duration of femtoseconds is emitted [28].
Although there is not yet a clear explanation of the sonoluminescence 
phenomenon and different theories have been proposed, the model presented 
here can give a simple and qualitative explanation of the phenomenon in 
terms of the dynamical Casimir effect. 
A pulse varying very quickly in time can describe very approximately
what happens when the bubble collapses and the periodicity can give an
idea how the bubble can emit light periodically. 
$$
$$
\vfill 
\eject
\par
\noindent
\bf 4. The static Casimir effect and thermodynamic instability
\bigskip
\rm 
A property shared by many one loop quantum statistical thermodynamic 
computations is that a thermodynamic second law instability appears in 
the final answer. Perhaps the most commonly discussed example of this 
phenomenon occurs in black hole statistical thermodynamics [29,30]. The entropy 
and the energy of a stationary non-charged, non-rotating black hole 
having mass $M$ are given by 
$$
S=4\pi k_B\Big({GM^2\over \hbar c}\Big), \ \ \ \ \ E=mc^2 \eqno(29)
$$
where $G$ is Newton's gravitational coupling strength. The black hole 
temperature, defined as $T=c^2(\partial M/\partial S)$, is then 
determined by 
$$
M=\Big({\hbar c^3 \over 8\pi G k_BT}\Big). \eqno(30)
$$ 
The black hole heat capacity $C=c^2(\partial M/\partial T)$ is 
thereby negative, 
$$
C= c^2 {\partial M \over \partial T}= - 
{K_B \hbar c^5 \over 8 \pi G T^2} <0, \eqno(31)
$$
therefore the black hole is thermodynamically instable. An increase in its 
energy decreases its temperature. This fact is considered by some authors as a 
possible violation of the second law of thermodynamics.

That a one loop quantum gravity calculation, i.e. the gravitational 
Casimir effect, produces thermodynamical instability is (perhaps) 
not very surprising. 
Even at the Newtonian theoretical level, the long 
range gravitational attraction upsets the usual  
convexity conditions otherwise present in the thermodynamic 
limit of infinite size. 

Our purpose is to show  that 
the electrodynamic Casimir effect can also produce thermodynamic instability.
To see what is involved, suppose that a material is located inside a box
of volume $V=Az$ with a movable piston.
The free energy per unit area obeys 
$$
df=-sdT-Pdz, \eqno(32)
$$
where $S=As$ and $F=Af$.
If the system is thermodynamically stable then the isothermal 
compressibility must be non-negative 
$$
K_T=-\Big({1\over V}\Big)\Big({\partial V\over \partial P}\Big)_T =
-\Big({1\over z}\Big)\Big({\partial z\over \partial P}\Big)_T \geq 0.
\eqno(33)
$$

However the condition given by Eq. (33) is not satisfied for the case of the
electromagnetic vacuum, which is therefore thermodynamically unstable.

Let $f(z,T)$ be the free energy per unit area of the vacuum inside two  
perfectly conducting non-charged parallel plate
The ground state energy per unit area of this vacuum is given by 

$$
\epsilon (z)=\lim_{T\to 0}f(z,T)=-\Big({\pi^2\over 720}\Big)
\Big({\hbar c\over z^3}\Big), \eqno(34)
$$
yielding the pressure  
$$
P_0=\lim_{T\to 0}P(z,T)=-\Big({\pi^2\over 240}\Big)
\Big({\hbar c\over z^4}\Big). \eqno(35)
$$
The zero temperature compressibility then reads 
$$
K_0=\lim_{T\to 0}K_T=-\Big({60\over \pi^2 }\Big)
\Big({z^4 \over \hbar c }\Big)<0. \eqno(36)
$$
The negative compressibility ($K_0<0$) in Eq.(36) violates the 
stability condition, as written in Eq.(33). 

The notion of negative compressibility matter is as old as the 
van der Waals approximation to the equations of state of a material [31]. 
 It has always been stated 
that such equations of state require supplementary conditions such as 
equal area constructions and so forth (see for example Ref. [31] pp 257-62). 
Furthermore, the second 
thermodynamic law has been thought to put a complete and total 
veto on observing the totally unstable part of the van der Waals 
curve; i.e. $K_T<0$ exists formally in the approximation, but is 
strictly forbidden from observation.   

So now we have a paradox, and perhaps an interesting energy source. 
For the Casimir force, and even for Coulomb's law, the regime  
$K_T < 0$ may be asserted to be real.

To end in order to see under 
which condition  there is stability,  consider the vacuum fluid in
a cylinder separated by an internal movable piston from a given material.
Then the total free energy of the system at zero temperature is 
$$
f_{tot}=f(z,0)+f'(z-L,0),
\eqno(37)
$$
where $f(z-L)'$ is the free energy of the 
material and $L$ is the total length of the cylinder with $V=AL$.

The stability condition dictates that 
$$
{\partial ^2 f'_{tot} \over \partial z^2} \geq 0,
\eqno(38)
$$
and therefore
$$
{1 \over K_o} +{1 \over K_m} \geq 0,
\eqno(39)
$$
where $K_m$ is the compressibility of the material. Hence thermodynamic 
stability is reached  when
$$
K_m \geq |K_o|,
\eqno(40)
$$
$$ 
$$
\bf 5. Summary
\bigskip
\rm
We have shown that it is possible to create
photon radiation by modulating the vacuum between two perfectly conducting
plates when the distance between them is sufficiently small and the
modulation is very rapid in time.

Moreover we have considered the stability condition associated with
the electromagnetic vacuum between two perfectly conduction 
non-charged parallel plates, which can be thermodynamically unstable.
\vfill
\eject

\bigskip
 
\centerline{\bf References}
\medskip

\par \noindent
\noindent [1]  H.~B.~G.~Casimir, {\it Pro. Kon. Ned. Akad. Wet.} {\bf
51}, 793 (1948).\par

\noindent [2] Yu. S. Barash and V. F. Ginzburg, 
{\it Usp. Fiz. Nauk.} ${\bf 116}$, 5 (1975). [Sov. Phys. Usp. ${\bf 18}$, 
305 (1975)].\par 

\noindent [3] G.  Plunien, B. Muller, and  W. Greiner   
1986 {\it Phys. Rep.} {\bf 134}, 87.\par   

\noindent [4]  P. W.  Milonni  
{\it The Quantum Vacuum} (London: Academic Press, 1994).\par

\noindent [5] V. M.   Mostepanenko   and N. N.  Trunov   
{\it The Casimir Effect and its Applications} \
 (Oxford: Clarendon Press, 1997). \par

\noindent [6]  J.~Schwinger, {\it Lett. Math. Phys.} 
{\bf 24}, 59 (1992); ibid {\bf 24}, 227 (1992); 
{\it Proc. Natl. Acad. Sci.} {\bf 89}, 4091 
(1992);  {\bf 89}, 11118 (1992); ibid {\bf 90}, 958  
 (1993); ibid {\bf 90}, 2105 (1993); ibid 
{\bf 90}, 4505 (1993); ibid {\bf 90}, 7285 (1993). \par

\noindent [7]   E. Sassaroli, {\it Photon and Fermion Pair Production
through the Modulation of the QED Vacuum}, Ph.D thesis, December 1993;   
A.~Widom, E.~Sassaroli and Y.~Srivastava, {\it Can. J. Phys.}
{\bf 71}, 168 (1993); E. Sassaroli, Y.~N. Srivastava, 
and A.~Widom, \it Phys. Rev. A \bf 50 \rm , 
1027 (1994);   \it Nucl.
Phys. B (Proc. Suppl.) \rm  ${\bf 33C}$, 209 (1993).\par

\noindent [8] C. K. Law, {\it Phys. Rev. Lett} {\bf 73}, 1931 (1994).

\noindent [9] V. V. Dodonov, {\it Phys. Lett.} A {\bf 207}, 126 (1995).

\noindent [10] P. A. Maia Neto and L. A. S. Machado, {\it Braz. J. Phys.}
{\bf 25}, 324 (1995).

\noindent [11] Kimball A. Milton, hep-ph/9510091; K. A. Milton, J. Ng,
hep-ph/9707122.

\noindent [12] V. I. Man'ko hep-ph/9502024.

\noindent [13] O. Meplan and C. Gignoux, {\it Phys. Rev. Lett}
{\bf 76}, 408 (1996).

\noindent [14] C. Eberlein, {\it Phys. Rev. Lett.} {\bf 76}, 3842 (1996), 
{\it Phys. Rev. A}
{\bf 53}, 2772 (1996).

\noindent [15] A. Lambrecht, M.-T. Jaekel, and S. Reynaud, 
{\it Phys. Rev. Lett.}{\bf 77}, 615 (1996).

\noindent [16] P. Davis, {\it Nature} {\bf 382}, 761 (1996).

\noindent [17] P. Knight, {\it Nature} {\bf 381}, 736 (1996).

\noindent [18] Alan Chados, hep-ph/9604368.

\noindent [19] C. E. Carlson, C. Molina-Paris, J. Perez-Mercader, and 
M. Visser, {\it Phys. Lett B} {\bf 395}. 76 (1997); {\it Phys. Rev. D}
{\bf 53}, 2772 (1996). 

\noindent [20] Jeong-Young Ji, Hyun-Hee, Jong-Woong Park, Kwang-Sup Soh,
hep-ph/9706007.

\noindent [21] Diego. A. R. Dalvit and Francisco D. Mazzitelli, 
hep-ph/9710048.

\noindent [22] R. Golestanian, M. Kardar, quant-ph/9802017.

\noindent [23] V. V. Nesterenko and I.G. Pirozhenko, JETP Lett. 
{\bf 67} 445 (1998).  

\noindent [24] S. Liberati, F. Belgiorno, Matt Visser, D. W. Sciama, 
hep-ph/9805031.

\noindent [25] A. Widom, E. Sassaroli, Y. N. Srivastava, and 
J. Swain, quant-ph/9803013.

\noindent [26] S. A. Fulling and P. C. W.  Davies, {\it Proc. R. Soc. London 
A} \bf 248  \rm , 393 (1976); S. A. Fulling, \it Aspects of Quantum Field 
Theory  in Curved Space-time \rm (Cambridge University Press 1989).\par

\noindent [27]  L.~S.~Brown and L.~J.~Carson, 
{\it Phys. Rev.} {\bf A20}, 2486 (1979). \par

\noindent [28] S. L. Putterman, {\it Scientific American}, February 1995. \par 

\noindent [29] Bekenstein J. D. 1973 {\it Phys. Rev. D} {\bf 7}, 2333.
\par 

\noindent [30] Hawking S. W. 1975 {\it Comm. Math. Phys.},
{\bf 43}, 199.\par

\noindent [31] Landau L. D. and Lifshitz E. M. 1994 
{\it Statistical Physics I} (Oxford: Pergamon Press).

\par

\bye